# Categorization of an emerging discipline in the world publication system (SCOPUS): E-learning

Gerardo Tibaná-Herrera[*], María Teresa Fernández-Bajón[†], Félix de Moya-Anegón[‡]

## ABSTRACT

E-learning has been continuously present in current educational discourse, thanks to technological advances, learning methodologies and public or organizational policies, among other factors. However, despite its boom and dominance in various subject areas, this thematic does not yet exist in the world system of publications. Therefore, works in this thematic end up being published under related categories, particularly Education or categories within subject area Computer Science, thus fragmenting and make invisible the existing knowledge. This work is based on the hypothesis that the scientific communication of e-learning has a sufficient degree of cohesion to be considered as a thematic category in itself. From a bibliometric approach, its scientific production was analyzed, obtaining the bibliographic data of SCOPUS and SCImago, selecting its main descriptors and generating visualizations through VOSViewer with the mapping overlay technique, to represent its set and proximity. As a result, it was determined that a set of 219 publications show a high bibliometric interrelation among its articles and these are presented transversely between the social sciences, computer science and health. This set serves as a channel of scientific communication and structure of knowledge on the thematic and can therefore be considered as the basis for establishing the "e-learning" thematic category in the world system of scientific publications, contributing to the consolidation of the discipline, to its access and development by researchers.

[*] Complutense University of Madrid. PhD Library Science Program. gtibana@ucm.es. SCImago Research Group. gtibana@gmail.com
[†] Complutense University of Madrid. Library Science Faculty. mfernandez@ucm.es
[‡] SCImago Research Group. felix.moya@scimago.es

# INTRODUCTION

E-learning, short for "electronic learning", arises as a name for the advances that have been made in education through the use of Information and Communication Technologies (ICT) and in particular the internet. Multiple definitions have emerged over the past few decades, it has been considered as a "new" form of learning (Nicholson, 2007) that uses the strengths of the Internet to provide synchronous or asynchronous interaction, teaching materials, and personalized programs to various communities. Even before the existence of the internet, interest was focused on the support that computer equipment and networks could offer to teachers and students, enriching education with technological findings, as told by Fuller (1962). Then, the interest turn towards the educational, on the one hand concerning learning (Stockley, 2006) and on the other hand to the understanding of this new modality as an evolution of the distance modality (Tick, 2006). Over time, it has been given a highly revolutionary character, granting it the capacity to transform education through the production, participation and consumption of content in various formats (Abeliuk, 2015). This same character has led to it being a focus of multiple views. UNESCO defines it as a fundamental element in 21st century education that contributes to the construction and participation of the knowledge society (UNESCO, 2013); is also considered as an object of transversal study for the development of the networked society (Guerrero, 2010); is also seen as a provider of tools and channels to access a new experiences, scenarios and information mediated by ICT, innovating in teaching-learning processes, research, extension (Freire, 2008) and other institutional practices, as described Conole y Oliver (2006), who also defines it as a field of research, with complex thematics, multiple tensions and rapid movement. Finally, Chiang determines on his work that e-learning research is expanding significantly (Chiang, Kuo , & Yang, 2010).

From the bibliometric point of view, one can learn a great deal by analyzing research manifested in scientific publications (journals and conference proceedings), as Taylor demonstrated at the beginning of the millennium (Taylor, 2001). Several studies have been carried out in educational sciences (Lee, Wu, & Tsai, 2009) (Diem & Wolter, 2013) and also from technology (Hsiao, Tang, & Liu, 2015). However, only few and limited studies have been conducted to know the scientific production in e-learning. One of the first was the work of Shih, Feng and Tsai (2008), based on 5 journals, highlighting the trends in "Instructional Approaches", "Learning Environments" and "Meta Cognition." A similar approach was taken by Maurer and Khan (2010) when analyzing five journals and two conferences proceedings, identifying 14 trends and 150 concept clusters. At the same time, a broader study of the Ed-Media journal archive from 2003 to 2008 (Khan, Ebner, & Maurer, 2009) was also conducted, finding problems with the ambiguity of terms, institutions and authors, emphasizing the need to create an appropriate and comprehensive thematic category. Chiang, Kuo and Yang (2010) found that the main applications of e-learning were presented mainly in Education and Computer Science and to a lesser extent in Medical Education, Information Sciences and Documentation, among other interdisciplinary areas. They used an empirical consultation of the terms "e-learning", "distance learning" and "electronic learning", focusing their analysis on 7 journals.

Changing the focus of journals and conferences proceedings to databases and indexes, Liu, Wu, and Chen (2013) took WoS and ScienceDirect to identify Learning Technologies trends in special education. González, Saroil and Sánchez (2015) analyzed the scientific production in Latin America with the SciELO database, putting on the table the growing thematic linkage of e-learning with the areas of Education (like Chiang, Kuo and Yang (2010)) and of Health Sciences, also highlighting its multidisciplinary nature. Finally, Tai, Lee and Lee (2013), using the SSCI and SCI, analyzed the citation of journals in the period 2003-2012, finding a diversity of tendencies in multiple areas of knowledge.

These studies show the increase in the scientific production of e-learning over time, along with its trends, as technology and educational practices bring innovations. However, they contribute to the exploration of the thematic rather than its appropriate categorical definition (Khan, Ebner, & Maurer, 2009). Therefore, we have for many reasons a large gap in the analysis of the scientific domain in this discipline, either because it is a very recent field (compared to other disciplines), or because it doesn't have a global panorama of contrast to the recognition of that category, or because changes in intellectual environments occur so rapidly and cover new conceptions of industry, education, and politics, or because most of today's scientific advances no longer align with disciplinary boundaries (Rafols, Porter, & Leydesdorff, 2010).

So how to know if e-learning is a discipline in itself? Well, this study seeks to answer this question from a bibliometric approach with the support of other visualization techniques, beyond the graphs of the indicators.

The combination of bibliometrics with visualization techniques to analyze and/or define emerging disciplines has already been used in other cases, such as Nanoscience and Nanotechnology (Muñoz, Vargas, Chinchilla, Gómez , & Moya-Anegón, 2011), where journals and conference proceedings were analyzed, consolidating the identity of the scientific discipline with a high degree of transversality and without defined limits; the case of Environment Management Accounting (Schaltegger, Gibassier, & Zvezdov, 2013), whose results show an incipient development of the thematic; the case of altimetry (González, Pacheco, & Arencibia, 2016), which used the terms that in the opinion of the authors, better define the thematic, identifying the research trends that characterize it.

Other studies have used bibliometrics to identify scientific publications with the highest impact in a certain discipline, for example in the field of information systems (Chan, Guness, & Kim, 2015), information science (Waltman, van Eck, & Noyons, 2010), environmental social responsibility (Valenzuela, Linares, & Suárez, 2015), renewable energy, sustainability and environment (Fernández, Bote, & Moya-Anegón, 2013), business and technical communication (Lowry, Humpherys, Malwitz, & Nix, 2007), neurosurgery (Madhugiri, Ambekar, Strom, & Nanda, 2013), among others.

As for bibliometric analysis, co-citation of classes or categories can be used to construct maps of large scientific domains (Moya-Anegón, et al., 2004), this may serve, as mentioned above, for the establishment of a global panorama of contrast on which different scientific fields can be recognized and their internal dynamics and cognitive structure understood (Cobo, López, Herrera, & Herrera, 2011), either as a field of research already consolidated or as an emerging discipline. In Guzman's words, "we can say that the analysis of information with maps of science, supported by metric studies of information, allows us to graphically represent the relationships between documents that the disciplines or specific scientific fields publish. These show the sub-areas of research in which the discipline has been focused over the years in order to identify, analyze and visualize the intellectual structure, as well as the temporal evolution in which the disciplines are being developed." (Guzmán & Trujillo, 2013).

Rafols, Porter, and Leydesdorff (2010) developed a method for visually locating research bodies within science. By overlaying maps of science we can investigate the increase in the scientific development of disciplines and organizations that do not fit into the traditional disciplinary categories, this is achieved thanks to the existence or construction of a stable corpus over which another smaller corpus can be superimposed, producing intuitive comparisons, with greater interpretation, and with the potential for use in scientific analysis and for comparative purposes (Boyack, 2008). Following the method, these maps are matrices of similarity measures, calculated from the correlation between information items present in the structure of scientific communication, in other words, show the disciplinary structure of the sciences in terms of publications. The stable or base map is constructed with bibliographic data of a database that has a defined categorization of the sciences. The analysis performed on the overlap will be conditioned by the size of the data selected for it.

An example of the use of the mapping overlay technique was developed and published by the SCImago[4] research group in its work on the graphical interface of SCImago Journal and Country Rank[5] (Hassan, Guerrero, & Moya-Anegón, 2014) in which through a freely accessible web platform, the presence of SCOPUS publications in different scientific domains can be analyzed, as well as the global distribution of the scientific output performance of different regions and countries. This tool also allows seeing the thematic categories with which the scientific publications have been previously related, both the traditional knowledge areas and the research frontiers.

---

[4] www.scimagolab.com
[5] www.scimagojr.com

Given the need for a database that could represent the global scientific publication system, we used SCOPUS[6] as a data source because of its disciplinary coverage, as did Leydesdorff, Moya-Anegón and Guerrero (2010), to carry out the comparative study between JCR and later to measure the interdisciplinarity of this database (Leydesdorff, Moya-Anegón, & Guerrero, 2015).

As for the visualization technique, there are multiple methods and tools for visualizing bibliometric networks, such as distance-based, graph-based or time-based (van Eck & Waltman, Visualizing bibliometric networks, 2014) (Small, 2006). Mapping and clustering are also used to respond to concerns about the main fields of research in a scientific domain, the relationship between research fields and the evolution of the domain over time. As a tool, Leydesdorff, Moya-Anegón and Guerrero (2015) in their journal mapping work showed how VOSViewer[7] assure the comprehensive visualization of node labels on the map and how the stress minimization technique such as multidimensional scaling (MDS) facilitates its visual.

## MATERIALS AND METHODS

This study is based on the existence of scientific communications that contribute to the development of the thematic, under the criteria of academic peer review and deposit in databases of international publications. In order to identify such communications and analyze them, the following methodology is developed.

**Step 1. Definition of descriptors.** Establish a list of all those terms that describe the thematic from the term core "e-learning." This is achieved through the bibliometric analysis of articles that in their title, abstract and keywords include the term core. It begins with the definition of the type of information, in this case the primary literature is considered as the main and most important reference of knowledge in the world scientific field (Fernández, Bote, & Moya-Anegón, 2013), for its scientific contributions and for receiving most of the citations. The source of information consulted was SCOPUS as the database that mostly indexes journals and conferences proceedings (Leydesdorff, Moya-Anegón, & Guerrero, 2010). The search results are refined by source type (journal and conference proceedings), by primary literature (article, conference paper and review), then the time period for the analysis is selected (2012-2014) and the language (English) and finally, the documents (2000) are selected on which to perform the bibliometric analysis of the keywords.

---

[6] SCOPUS is the largest abstract and citation database of peer-reviewed literature: scientific journals, books and conference proceedings. A Elsevier's product. https://www.elsevier.com/solutions/scopus
[7] www.vosviewer.com

A bibliometric analysis based on the co-occurrence of keywords was performed, with the aim of determining primary descriptors that are mainly present in articles, their relationships and relevance by means of the technique of Visualization of Similarities (VoS) (Waltman, van Eck, & Noyons, 2010). This technique, as shown by Cobo, Lopez and Herrera (2011), provide a very accurate look at how a document corpus is described and linked.

**Step 2. Correspondence of publications and descriptors.** Build a matrix of articles volume for each term and each publication indexed in the database. Using the same selection criteria described in the previous step, a query is made to the database for each of the descriptors that have appeared thus determining the number of articles of each descriptor. It is also necessary to include new descriptors as result of linguistic similarities and/or acronyms or abbreviations used in natural language, for example, when including the keywords of an article, you can choose to use the descriptor *e-learning* or *elearning* (Chiang, Kuo , & Yang, 2010), or the acronym *ICT* to include the descriptor *Information and Communication Technologies*. These new descriptors, that reflect the same meaning as the one provided by the author, are called Secondary Descriptors and constitute new columns in the correspondence matrix. A final phase corresponds to the sum of the related articles with the primary and secondary descriptors of each term, assuming that the sum reflects unique works related by descriptors.

**Step 3. Percentage of participation in the thematic.** Determine the percentage of articles in the journal or conference proceeding that are related to the thematic during the time period established in the initial criteria. The total number of articles (TNA) of the journal is identified during the period of time, then the number of related articles (NRA) with the thematic is determined for each of the journals, this is done by taking the maximum number of articles by descriptor, considering that an article can be related to more than one descriptor. Then, the percentage of participation (PP) between these values is calculated:

$$PP = \frac{NRA}{TNA} \times 100$$

**Step 4. Cut-off point for inclusion of publications in the category.** The cut-off point should be determined on the PP from which the publications will be included for the thematic categorization. Previous studies on this classification task have been carried out on the basis of the distribution of publications between "pure", "hybrid" and "non-related" publications (Chan, Guness, & Kim, 2015) or on the determination of the core set of publications (Madhugiri, Ambekar, Strom, & Nanda, 2013). However, it was considered that this cut-off point must be established by identifying the maximum permissible error of the thematic relation of the publication. The higher the cut-off point the greater the precision in the selection of publications, although this precision means a reduced volume, and if not, a low cut-off point increases the error in the selection and its volume. Once the cut-off point is established, all publications that exceed this threshold will be considered in the categorization of the emerging discipline.

**Step 5. Analysis of the set of publications.** The degree of cohesion between publications is sought. The selected journals are analyzed under a bibliometric view to determine if they represent the existence of a scientific community that communicates their knowledge through these channels, to recognize it as an emergent and distinctive scientific discipline that can be delimited as a cross-thematic category (Leydesdorff, Moya-Anegón, & Guerrero, 2015). In this study we will use the mapping overlay technique (Rafols, Porter, & Leydesdorff, 2010) that facilitates the exploration of the knowledge bases of an emerging discipline and its evolutionary dynamics both in terms of its internal cognitive coherence and diversity of their sources of knowledge with reference to disciplinary classifications. For this, it is necessary to have a base map on which to overlay a local map and thus make comparisons.

The base map will be a global map of science (Leydesdorff, Moya-Anegón, & Guerrero, 2015) that includes the total of journals and conference proceedings indexed in SCOPUS. The relationship degree of publications is established by the normalized value produced by the combination of cites, co-cites and coupling (Hassan, Guerrero, & Moya-Anegón, 2014). In addition, this analysis is enriched with the clustering performed by VOSViewer (van Eck & Waltman, 2010).

The local map that will be overlaid on the global map of science is the set of journals and conference proceedings selected in the previous step. This overlap will allow locating the discipline in the general topology of scientific knowledge and whether or not there is a cluster effect, which should be considered as evidence of the existence of a specific disciplinary field from the point of view of the scientific communication guidelines followed by researchers.

In summary, the methodology is based on the principle that a greater presence of field-specific descriptors in the items of an article is directly proportional to the number of interactions by citation, co-citation, and coupling of a publication with others that would form part of the discipline cluster.

## RESULTS AND DISCUSSION

The SCOPUS database was used for its coverage and peer review of indexed publications to extract all primary literature that includes the term e-learning in its title, abstract and keywords. The database was consulted with the query chart of Table 1, obtaining the metadata of the first 2000 publications with a total of 4521 keywords[8].

| Source type | Journal, Proceedings |
|---|---|
| Document type | Article, Conference paper, Conference review |
| Timespan | 2012, 2013, 2014 |
| Languaje | English |

**Table 1.** Query chart on SCOPUS

These keywords were later analyzed with the VOSViewer tool to establish their co-occurrence in the articles.

---

[8] Search made on January 23, 2017.

Under this technique, 51 primary descriptors were established and are listed in Figure 1, including the occurrences of the term core as well.

The secondary descriptors that complete the listing are: (52) elearning, (53) electronic learning, (54) Learning management system, (55) blearning, (56) blended learning, (57) mlearning, (58) mobile learning, (59) Information and communications technologies, (60) eassessment, (61) electronic assessment, (62) VLE, (63) Massive Open Online Courses, (64) PLE. These 64 terms constitute the base descriptors of the consultation of the articles, since they have been used to describe the scientific production around e-learning during the established period.

The correspondence matrix between descriptors and publications was elaborated on a total basis of 12,923 journals and conferences indexed in SCOPUS[9]. The PP participation percentage is shown in Figure 2.

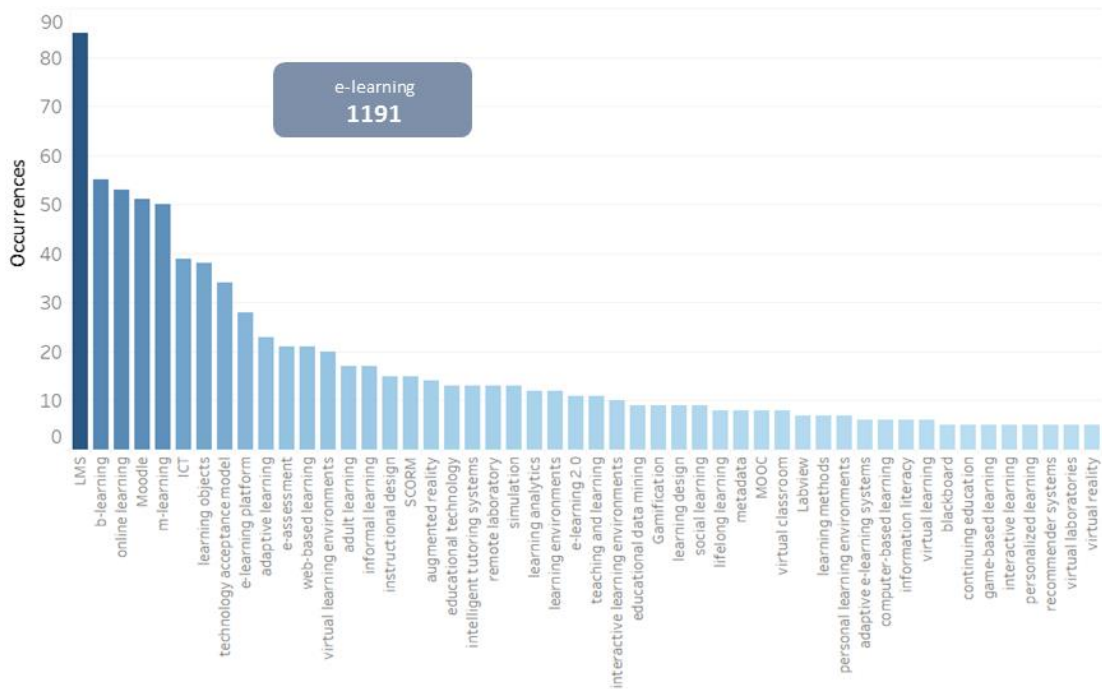

**Figure 1.** "E-learning" primary descriptors

---

[9] Search made on February 12, 2017.

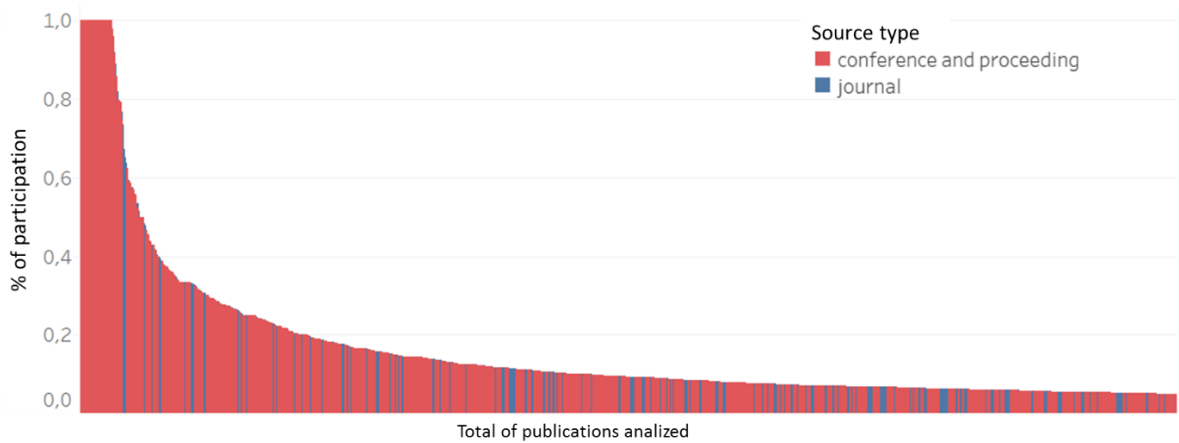

**Figure 2.** Percentage of participation (PP) of the term in journals and conferences

There are 3,680 publications of the total base that do not have any article related to any of the 64 descriptors. Of the remaining 9,243, 7,801 have a participation rate of less than 5%. The cut-off point setting was performed based on the analysis of the error in selection of publications, in this case, by percentage bands as shown in Table 2.

| Band | % of participation | Publications | Error | error % | Average Participation Percentage |
|---|---|---|---|---|---|
| 1 | 100 | 44 | 0 | 0% | 100% |
| 2 | > 95 | 45 | 0 | 0% | 100% |
| 3 | > 90 | 47 | 0 | 0% | 100% |
| 4 | > 85 | 49 | 0 | 0% | 99% |
| 5 | > 80 | 51 | 0 | 0% | 98% |
| 6 | > 75 | 57 | 0 | 0% | 96% |
| 7 | > 70 | 58 | 0 | 0% | 96% |
| 8 | > 65 | 60 | 0 | 0% | 95% |
| 9 | > 60 | 64 | 0 | 0% | 93% |
| 10 | > 55 | 74 | 0 | 0% | 88% |
| 11 | > 50 | 84 | 0 | 0% | 84% |
| 12 | > 45 | 91 | 0 | 0% | 81% |
| 13 | > 40 | 105 | 0 | 0% | 76% |
| 14 | > 35 | 125 | 0 | 0% | 70% |
| 15 | > 30 | 169 | 4 | 2% | 60% |
| 16 | > 25 | 230 | 11 | 5% | 51% |
| 17 | > 20 | 299 | 20 | 7% | 45% |

**Table 2.** Percentage bands for establishing the cut-off point.

As can be seen, a value less than 20% has an error greater than 7% and the average participation percentage is less than 50%. We consider that an average percentage of participation should be maintained above 50% for the publication be considered in the category, which is why the cut-off point is set at 25% (coinciding with the classification of pure and hybrid publications made by Chan, Guness and Kim (2015)) and excluding the 11 journals and congresses that are not related to the theme. With this, there are 82 journals and 137 congresses (Annex 1) to be analyzed in comparison to the global map of science.

The global science map was constructed using the VOSViewer from SCOPUS-indexed publications and the combined indicator (citations, co-citations, coupling) used by SCImago (Hassan, Guerrero, & Moya-Anegón, 2014), which guarantees to have normalized values for each of the publications to be visualized.

As can be seen in Figure 3, the map is composed of seven clusters, which in a clockwise and wide sense can be denominated as: Social Sciences (red), Psychology (clear cyan), Medicine (green), Health Sciences (purple), Life sciences (yellow), Physical sciences and Engineering (dark cyan) and Computer science (blue).

Figure 4 presents the overlap of the local map corresponding to the 219 selected publications (black color) on the global map of science, showing the distribution of publications.

Figure 5 clearly shows the cluster effect that shows a high interrelation of combined indicator (citations, co-citations, coupling). This cohesion is sufficient evidence, in terms of scientific communication, that there is a shared use of publications among researchers of this thematic and determines that e-learning is a distinctive scientific discipline. For there is a network of relationships and interactions that are established between researchers who share structures of thought, patterns of cooperation, language and forms of communication, as Hjorland and Albrechtsen (1995) put it in establishing that science must be evaluated Knowledge of the social practices of scientists.

There is also a core of publications within the cluster. An analysis of the 26 publications in the core shows that 77% are classified in the Education category, 50% in the Computer Science category and 23% in Documentation Sciences. However, the high cohesion that forms this core is because 65% of publications classified in Education are also classified in Computer Science.

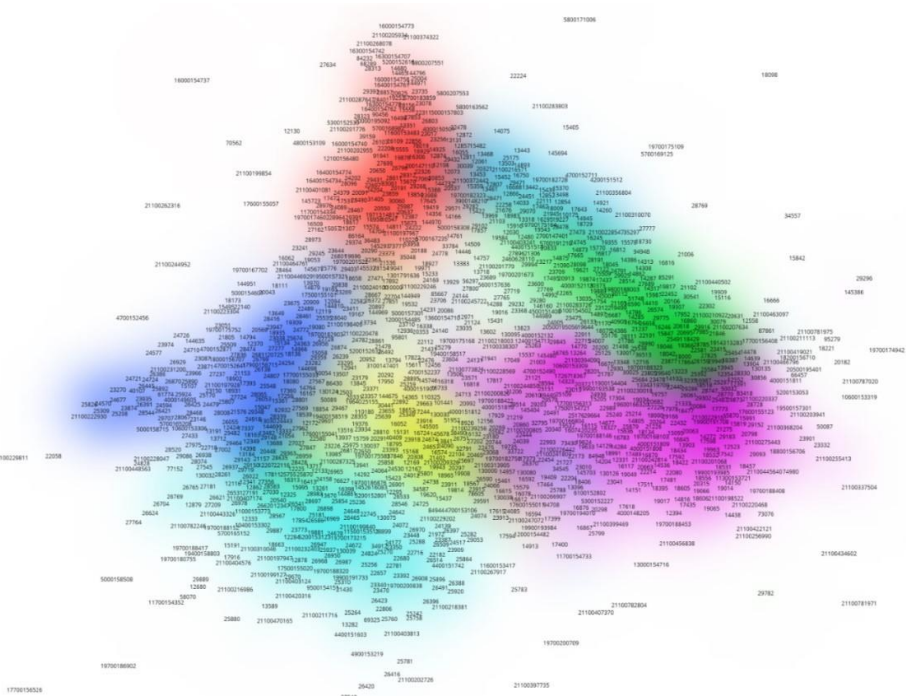

**Figure 3.** Global map of science based on SCOPUS and SCImago using VOSViewer with its density map setting (Source: self-made)

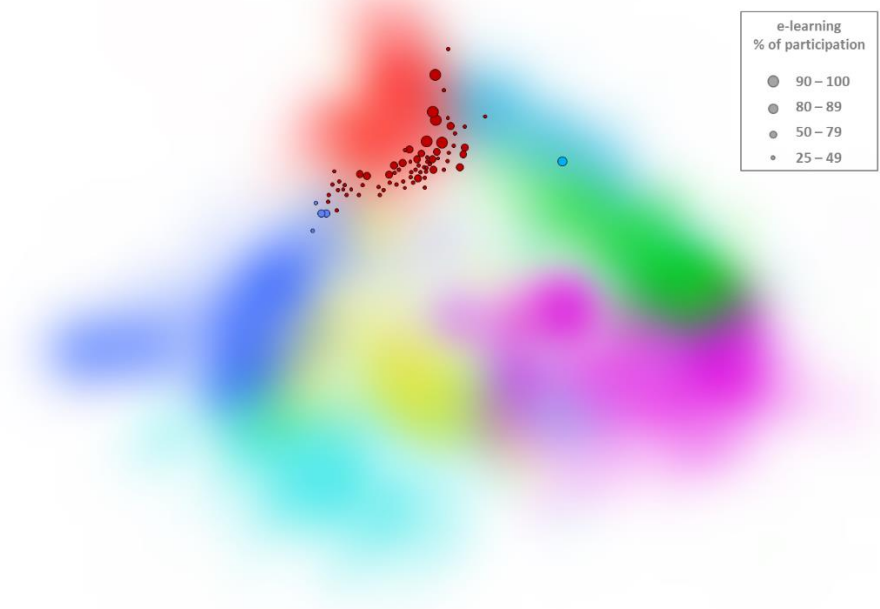

**Figure 4.** Distribution of publications related to the thematic, using the mapping technique with VOSViewer in its density map configuration. (Source: self-made)

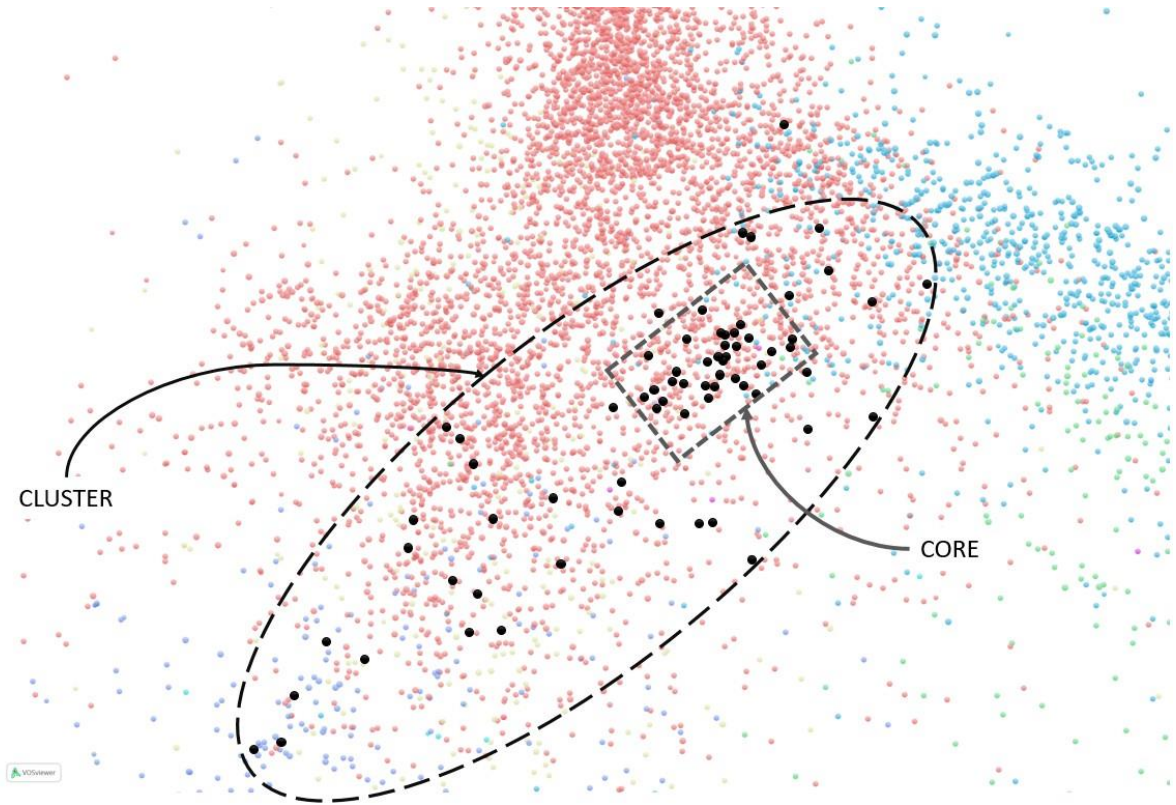

**Figure 5.** Approximation of the distribution of publications related to the thematic, by map overlay in VOSViewer in its configuration of network map without links. The size of the selected publications has been modified for visual purposes. (Source: self-made)

On the other hand, the cluster can be contrasted and validated by categorizing the selected publications into an existing indexing system, for example, SCImago Journal Rankings. Figure 6 is the result of an analysis of common categories among the selected publications, plotted using the NodeXL tool with the force-direct visualization algorithm of Fruchterman and Reingold (1991). As can be seen, the strongest relationships are between Computer Science and Social Sciences, and then between them and Engineering and Management Sciences.

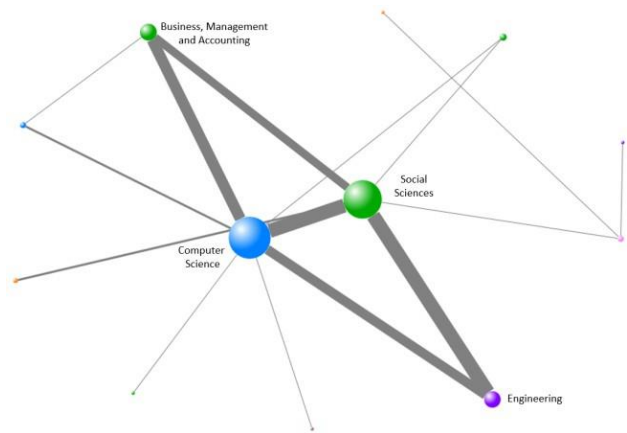

**Figure 6.** Relationship of thematic categories of the selected publications according to their classification in SCImagoJR. The colors of the categories of the same source were used (Source: self-made)

Although the ratio is inverse, a strong relationship between the Social Sciences and Computer Sciences is coherent and validates the behavior analysis of the core previously made.

In addition to the above, these findings diverge in part from the findings of Chiang, Kuo and Yang (2010) that e-learning is mainly present in Social Sciences and Computer Sciences.

This study demonstrates that the centrality of the e-learning thematic is in Social Sciences and in second order it is presented transversally towards Computer Sciences and Health Sciences.

## Discussion

There is no consensus in the scientific community about the descriptors of e-learning since they respond to different approaches given from pedagogy, technology, and organizations. For this reason, some descriptors considered important in the discourse may not have been included in this study, as well as other types of publications such as books, editorials, notes or surveys. The bibliometric analysis of the keywords provides an objective approximation in the construction of the set of descriptors, taking the denomination provided by the authors of the scientific production.

On the other hand, the categorization of scientific publications is an arduous and permanent task that is complex to manage from the publication systems, since a publication can vary over time in its central thematic and go to other research fronts without affecting its Initial categorization. This may be one reason why both the co-category analysis performed with the SCImagoJR and the previous categorization studies coincide only in the main areas, Social Sciences and Computer Science.

Likewise, this study can be updated under the same methodology as the corpus of articles and publications increases, generating new overlays and finally updating the thematic coverage.

## Possible applications

This categorization of e-learning is first, a guide for researchers who wish to know and contribute to the development and strengthening of the discipline, knowing the journals and congresses that comprise it, its impact and other bibliometric indicators. Secondly, it is an input for the development of new studies on the thematic, like georeferencing studies (Guerrero-Bote, Olmeda-Gómez, & Moya-Anegón, 2016), research of countries, institutions and authors and all kinds of bibliometric analysis. Finally, with this thematic categorization, the e-learning category can be included in databases (SCOPUS, WoS) and bibliometric analysis platforms (SCImago Journal Rank & Country Rank) to facilitate access and analysis of publications related to the thematic.

The methodology and tools presented in this study can be used in principle, for the analysis of any other scientific field or possible emergent discipline. However, it is necessary to consider the prerequisites to ensure the verification of the analysis under the global science landscape, these are, access to databases that represent global scientific knowledge or an approximation to it and, standard values for each of the publications to be analyzed.

# Conclusions

Using a combination of bibliometric indicators and analysis techniques, this study has categorized e-learning as an emerging discipline in the world system of scientific publications, consisting of 64 descriptors and 219 journals and congresses indexed by SCOPUS between 2012 and 2014.

According to other studies, the visualization analysis achieved by the map overlay technique clearly exposes a cluster effect of the global production of scientific knowledge in e-learning, that is to say, the existence of a concentration of scientific production with high degree of cohesion between the citation, co-citation and coupling indicators, which constitute a channel of scientific communication of the thematic within Social Sciences (Education) and with extensive and strong bibliometric links between Health Sciences (Psychology) and Computer Science, making it an emerging discipline of transversal character in the world system of publications.

This discipline must be analyzed from its internal structure (both cluster and core) to identify common principles, to define its nature, its detailed thematic correspondence and its main contributions and contributors.

The bibliometric indicators used in this study are an approximation the impact of publication in the scientific community and as such help to solve the problem of lack of consensus on the definition and description of the thematic, providing a set of descriptors that can be increased over time, by including annual scientific production in databases.

# Declarations

Availability of data and material

The data related to this research were obtained, on the one hand, from the access to SCOPUS and on the other, provided by SCImago Research Group. These are protected by licensing and copyright respectively.

## Competing interests

Not applicable.

## Funding

Not applicable.

## Authors' contributions

Primary author: Gerardo Tibaná-Herrera.

Analysis and review: María Teresa Fernández-Bajón, Félix de Moya-Anegón

## Acknowledgements

A SCImago Research Group por proveer los datos de citación de las publicaciones.

## Annexes

Annex 1 – List of publications considered for the categorization of the e-learning discipline

| SOURCE TITLE | TYPE | SOURCE TITLE | TYPE |
| --- | --- | --- | --- |
| 11th IEEE International Symposium on Mixed and Augmented Reality 2012 - Arts, Media, and Humanities Papers, ISMAR-AMH 2012 | conference and proceeding | Proceedings - 10th International Conference on Creating, Connecting and Collaborating through Computing, C5 2012 | conference and proceeding |
| 15th International Conference on Intelligent Games and Simulation, GAME-ON 2014 | conference and proceeding | Internet and Higher Education | journal |
| 2012 3rd International Workshop on Recommendation | conference and | Proceedings - 2014 International Conference of | conference and |
| Systems for Software Engineering, RSSE 2012 - Proceedings | proceeding | Educational Innovation Through Technology, EITT 2014 | proceeding |
| 2012 IEEE Symposium on E-Learning, E-Management and E-Services, IS3e 2012 | conference and proceeding | Proceedings of the 2014 IEEE International Conference on MOOCs, Innovation and Technology in Education, IEEE MITE 2014 | conference and proceeding |
| 2012 International Conference on Education and e-Learning Innovations, ICEELI 2012 | conference and proceeding | Reference Services Review | journal |
| 2012 International Conference on E-Learning and E-Technologies in Education, ICEEE 2012 | conference and proceeding | ITALICS Innovations in Teaching and Learning in Information and Computer Sciences | journal |
| 2013 2nd International Conference on E-Learning and E-Technologies in Education, ICEEE 2013 | conference and proceeding | Journal of Library and Information Services in Distance Learning | journal |
| 2013 IEEE Conference on e-Learning, e-Management and e-Services, IC3e 2013 | conference and proceeding | PATCH 2012 - Proceedings of the 2012 ACM Workshop on Personalized Access to Cultural Heritage, Co-located with ACM Multimedia 2012 | conference and proceeding |
| 2013 IEEE International Symposium on Mixed and Augmented Reality - Arts, Media, and Humanities, ISMAR-AMH 2013 | conference and proceeding | 22nd International Symposium on Human Factors in Telecommunication, HFT 2013 | conference and proceeding |
| 2013 IEEE International Symposium on Mixed and Augmented Reality, ISMAR 2013 | conference and proceeding | Revista Española de Lingüística Aplicada | journal |
| 3rd International Conference on Advances in New Technologies, Interactive Interfaces and Communicability, ADNTIIC 2012: Design, E-Commerce, E-Learning, E-Health, E-Tourism, Web 2.0 and Web 3.0 | conference and proceeding | 30th Annual conference on Australian Society for Computers in Learning in Tertiary Education, ASCILITE 2013 | conference and proceeding |
| 3rd International Conference on eLearning and eTeaching, ICeLeT 2012 | conference and proceeding | International Journal of Artificial Intelligence in Education | journal |
| 4th International Conference on e-Learning and e-Teaching, ICELET 2013 | conference and proceeding | International Journal of Electronic Finance | journal |
| IADIS International Conference on Cognition and Exploratory Learning in Digital Age, CELDA 2012 | conference and proceeding | 2012 2nd International Workshop on Games and Software Engineering: Realizing User Engagement with Game Engineering Techniques, GAS 2012 - Proceedings | conference and proceeding |
| IADIS International Conference on Cognition and Exploratory Learning in Digital Age, CELDA 2013 | conference and proceeding | Doctoral Student Consortia - Proceedings of the 21st International Conference on Computers in Education, ICCE 2013 | conference and proceeding |
| IASTED Multiconferences - Proceedings of the IASTED International Conference on Web-Based Education, WBE 2013 | conference and proceeding | International Information and Library Review | journal |
| IC3e 2014 - 2014 IEEE Conference on e-Learning, e-Management and e-Services | conference and proceeding | Journal of Research and Practice in Information Technology | journal |
| ISMAR 2012 - 11th IEEE International Symposium on Mixed and Augmented Reality 2012, Science and Technology Papers | conference and proceeding | International Journal of Learning Technology | journal |
| ISMAR 2014 - IEEE International Symposium on Mixed and Augmented Reality - Media, Arts, Social Science, Humanities and Design 2014, Proceedings | conference and proceeding | Proceedings of the European Conference on e-Government, ECEG | conference and proceeding |
| MARS 2014 - Proceedings of the 2014 Workshop for Mobile Augmented Reality and Robotic Technology-Based Systems, Co-located with MobiSys 2014 | conference and proceeding | British Journal of Educational Technology | journal |
| Proceedings - 2012 14th Symposium on Virtual and Augmented Reality, SVR 2012 | conference and proceeding | 2014 International Symposium on Computers in Education, SIIE 2014 | conference and proceeding |
| Proceedings - 2012 6th IEEE International Conference on E-Learning in Industrial Electronics, ICELIE 2012 | conference and proceeding | Proceedings - 2012 IEEE International Conference on Technology Enhanced Education, ICTEE 2012 | conference and proceeding |
| Proceedings - 2012 Technologies Applied to Electronics Teaching, TAEE 2012 | conference and proceeding | CEBE Transactions | journal |
| Proceedings - 2013 15th Symposium on Virtual and Augmented Reality, SVR 2013 | conference and proceeding | Mining the Digital Information Networks - Proceedings of the 17th International Conference on Electronic Publishing, ELPUB 2013 | conference and proceeding |
| Proceedings - 2013 4th International Conference on e-Learning Best Practices in Management, Design and Development of e-Courses: Standards of Excellence and Creativity, ECONF 2013 | conference and proceeding | Proceedings of 23rd International Conference on Artificial Reality and Telexistence, ICAT 2013 | conference and proceeding |
| Proceedings - 2013 7th IEEE International Conference on e-Learning in Industrial Electronics, ICELIE 2013 | conference and proceeding | 2012 International Symposium on Computers in Education, SIIE 2012 | conference and proceeding |
| Proceedings - 2014 16th Symposium on Virtual and Augmented Reality, SVR 2014 | conference and proceeding | Proceedings of the AIS SIG-ED IAIM 2013 Conference | conference and proceeding |
| Proceedings of the 2013 IEEE International Conference in MOOC, Innovation and Technology in Education, MITE 2013 | conference and proceeding | ASCILITE 2012 - Annual conference of the Australian Society for Computers in Tertiary Education | conference and proceeding |
| Proceedings of the European Conference on e-Learning, ECEL | conference and proceeding | International Telecommunication Union - Proceedings of the 2013 ITU Kaleidoscope Academic Conference: Building Sustainable Communities, K 2013 | conference and proceeding |
| Proceedings of the IADIS International Conference e-Learning 2012 | conference and proceeding | College and Undergraduate Libraries | journal |
| Proceedings of the IASTED International Conference on Computers and Advanced Technology in Education, CATE 2014 | conference and proceeding | Journal of Asynchronous Learning Network | journal |
| Proceedings of the International Conference e-Learning 2013 | conference and proceeding | CSEDU 2012 - Proceedings of the 4th International Conference on Computer Supported Education | conference and proceeding |
| Proceedings of the International Conference e-Learning 2014 - Part of the Multi Conference on Computer Science and Information Systems, MCCSIS 2014 | conference and proceeding | Proceedings of the 1st International Workshop on Interaction Design in Educational Environments, IDEE 2012, in Conjunction with ICEIS 2012 | conference and proceeding |
| Proceedings of the International Conference on Dublin Core and Metadata Applications | conference and proceeding | International Journal of Lifelong Education | journal |
| RecSys 2013 - Proceedings of the 7th ACM Conference on Recommender Systems | conference and proceeding | Proceedings of the 8th International Conference on Standardization and Innovation in Information Technology, SIIT 2013 | conference and proceeding |
| RecSys'12 - Proceedings of the 6th ACM Conference on Recommender Systems | conference and proceeding | IDIMT 2014: Networking Societies - Cooperation and Conflict, 22nd Interdisciplinary Information Management Talks | conference and proceeding |
| UXeLATE 2012 - Proceedings of the 2012 ACM International Workshop on User Experience in e-Learning and Augmented Technologies in Education, Co-located with ACM Multimedia 2012 | conference and proceeding | Encyclopaideia | journal |
| Adult Education Quarterly | journal | ICETA 2012 - 10th IEEE International Conference on Emerging eLearning Technologies and Applications, Proceedings | conference and proceeding |
| Educational Technology and Society | journal | 2014 IST-Africa Conference and Exhibition, IST-Africa 2014 | conference and proceeding |
| Interactive Technology and Smart Education | journal | 2012 International Conference on Information Technology Based Higher Education and Training, ITHET 2012 | conference and proceeding |
| International Journal of E-Adoption | journal | Transforming Government: People, Process and Policy | journal |
| Journal of Information Literacy | journal | Journal of Global Information Technology Management | journal |
| Turkish Online Journal of Educational Technology | journal | Journal of Information Technology Education: Research | journal |
| 22nd Italian Symposium on Advanced Database Systems, SEBD 2014 | conference and proceeding | Journal of Interactive Online Learning | journal |
| Proceedings of the International Conference on e-Learning, ICEL | conference and proceeding | Medical Physiology Online | journal |
| Journal of Continuing Education in the Health Professions | journal | Pakistan Journal of Library and Information Science | journal |
| eLmL - International Conference on Mobile, Hybrid, and On-line Learning | conference and proceeding | 2013 1st Workshop on Virtual and Augmented Assistive Technology, VAAT 2013; Co-located with the 2013 Virtual Reality Conference - Proceedings | conference and proceeding |
| International Journal of Mobile and Blended Learning | journal | ITiCSE-WGR 2014 - Working Group Reports of the 2014 Innovation and Technology in Computer Science Education Conference | conference and proceeding |
| L@S 2014 - Proceedings of the 1st ACM Conference on Learning at Scale | conference and proceeding | MCS 2013 - Proceedings of the 4th ACM Workshop on Mobile Cloud Computing and Services | conference and proceeding |
| Communications in Information Literacy | journal | Proceedings - 2014 International Conference on Interactive Technologies and Games, iTAG 2014 | conference and proceeding |
| Information Technology for Development | journal | Proceedings - 6th International Conference on Multimedia, Computer Graphics and Broadcasting, MulGraB 2014 | conference and proceeding |
| RecSys 2014 - Proceedings of the 8th ACM Conference on Recommender Systems | conference and proceeding | Proceedings of the 3rd Narrative and Hypertext Workshop Held at the ACM Conference on Hypertext and Social Media, NHT 2013 | conference and proceeding |
| International Journal of Mobile Learning and Organisation | journal | Proceedings of the 4th International Workshop on Modeling Social Media: Mining, Modeling and Recommending 'Things' in Social Media, MSM 2013 | conference and proceeding |
| ICT for Sustainability 2014, ICT4S 2014 | conference and proceeding | Proceedings of the 7th European Workshop on System Security, EuroSec 2014 | conference and proceeding |
| 2013 10th International Conference on Remote Engineering and Virtual Instrumentation, REV 2013 | conference and proceeding | Proceedings of the IADIS International Conference ICT, Society and Human Beings 2012, Proceedings of the IADIS International Conference e-Commerce 2012 | conference and proceeding |
| Computers and Education | journal | Electronic Journal of e-Learning | journal |
| MLA 2014 - Proceedings of the 2014 ACM Multimodal Learning Analytics Workshop and Grand Challenge, Co-located with ICMI 2014 | conference and proceeding | International Journal of Information and Communication Technology Education | journal |
| Learning Environments Research | journal | Distance Education | journal |
| International Journal of Virtual and Personal Learning | journal | Journal of Business and Finance Librarianship | journal |
| Environments | | | |
| Australian Educational Computing | journal | Revista Iberoamericana de Tecnologias del Aprendizaje | journal |
| Journal of Library Metadata | journal | International Journal of Web-Based Learning and Teaching Technologies | journal |
| Australasian Journal of Educational Technology | journal | IMSCI 2012 - 6th International Multi-Conference on Society, Cybernetics and Informatics, Proceedings | conference and proceeding |
| Doctoral Student Consortium Proceedings of the 20th International Conference on Computers in Education, ICCE 2012 | conference and proceeding | 11th International Conference on Cognition and Exploratory Learning in Digital Age, CELDA 2014 | conference and proceeding |
| Proceedings - 2012 International Symposium on Ubiquitous Virtual Reality, ISUVR 2012 | conference and proceeding | Internet Reference Services Quarterly | journal |
| Electronic Journal of Information Systems in Developing Countries | journal | CSEDU 2014 - Proceedings of the 6th International Conference on Computer Supported Education | conference and proceeding |
| 7th European Conference on Games Based Learning, ECGBL 2013 | conference and proceeding | Proceedings - 2013 2nd Experiment@ International Conference, exp.at 2013 | conference and proceeding |
| Proceedings of the European Conference on Games-based Learning | conference and proceeding | Community and Junior College Libraries | journal |
| 2012 9th International Conference on Remote Engineering and Virtual Instrumentation, REV 2012 | conference and proceeding | International Journal of Electronic Government Research | journal |
| Proceedings 2012 17th IEEE International Conference on Wireless, Mobile and Ubiquitous Technology in Education, WMUTE 2012 | conference and proceeding | Proceedings - 2012 IEEE 4th International Conference on Technology for Education, T4E 2012 | conference and proceeding |
| International Review of Education | journal | Knowledge Management and E-Learning | journal |
| International Journal of Continuing Engineering Education and Life-Long Learning | journal | Multicultural Education and Technology Journal | journal |
| Proceedings of 2014 11th International Conference on Remote Engineering and Virtual Instrumentation, REV 2014 | conference and proceeding | 2014 6th International Conference on Games and Virtual Worlds for Serious Applications, VS-GAMES 2014 | conference and proceeding |
| Proceedings of the 2012 IEEE Conference on Technology and Society in Asia, T and SA 2012 | conference and proceeding | International Journal of Online Engineering | journal |
| RINA, Royal Institution of Naval Architects - Education and Professional Development of Engineers in the Maritime Industry, Papers | conference and proceeding | SIGGRAPH Asia 2012 Symposium on Apps, SA 2012 | conference and proceeding |
| Journal of Educators Online | journal | Work-in-Progress Poster (WIPP) Proceedings of the 21st International Conference on Computers | conference and proceeding |
| | | in Education, ICCE 2013 | |
| International Journal of Distance Education Technologies | journal | International Journal of Technologies in Learning | journal |
| International Journal of Interactive Mobile Technologies | journal | Proceedings 2012 4th IEEE International Conference on Digital Game and Intelligent Toy Enhanced Learning, DIGITEL 2012 | conference and proceeding |
| IDIMT 2012 - ICT Support for Complex Systems, 20th Interdisciplinary Information Management Talks | conference and proceeding | 2012 International Conference on Interactive Mobile and Computer Aided Learning, IMCL 2012 | conference and proceeding |
| International Journal of Game-Based Learning | journal | 2014 1st International Conference on eDemocracy and eGovernment, ICEDEG 2014 | conference and proceeding |
| European Conference on Optical Communication, ECOC | conference and proceeding | Proceedings of the 2014 ITU Kaleidoscope Academic Conference: Living in a Converged World - Impossible Without Standards?, K 2014 | conference and proceeding |
| 27th Bled eConference: eEcosystems - Proceedings | conference and proceeding | SIGGRAPH Asia 2013 Symposium on Mobile Graphics and Interactive Applications, SA 2013 | conference and proceeding |
| LocalPeMA'12 - Proceedings of the 2012 RecSys Workshop on Personalizing the Local Mobile Experience | conference and proceeding | Proceedings of XI Tecnologias Aplicadas a la Ensenanza de la Electronica (Technologies Applied to Electronics Teaching), TAEE 2014 | conference and proceeding |
| Proceedings of PDSW 2014: 9th Parallel Data Storage Workshop - Held in Conjunction with SC 2014: The International Conference for High Performance Computing, Networking, Storage and Analysis | conference and proceeding | CSEDU 2013 - Proceedings of the 5th International Conference on Computer Supported Education | conference and proceeding |
| International Journal of Metadata, Semantics and Ontologies | journal | Proceedings of the 12th IEEE International Conference on Advanced Learning Technologies, ICALT 2012 | conference and proceeding |
| 2013 IST-Africa Conference and Exhibition, IST-Africa 2013 | conference and proceeding | Journal of Computer Assisted Learning | journal |
| Interactive Learning Environments | journal | DCNET 2014 - Proceedings of the 5th International Conference on Data Communication Networking, Part of ICETE 2014 - 11th International Joint Conference on e-Business and Telecommunications | conference and proceeding |
| Education and Information Technologies | journal | Proceedings - 2013 IEEE 5th International Conference on Technology for Education, T4E 2013 | conference and proceeding |
| International Journal of Emerging Technologies in Learning | journal | Ubiquitous Learning | journal |
| 2012 e-Leadership Conference on Sustainable e-Government and e- Business Innovations, E-LEADERSHIP 2012 | conference and proceeding | ICETA 2013 - 11th IEEE International Conference on Emerging eLearning Technologies and Applications, Proceedings | conference and proceeding |
| Doctoral Student Consortium (DSC) - Proceedings of the 22nd International Conference on Computers in Education, ICCE 2014 | conference and proceeding | Management and Technology in Knowledge, Service, Tourism and Hospitality - Proc. of the Annual Int. Conf. on Management and Technology in Knowledge, Service, Tourism and Hospitality 2013, SERVE 2013 | conference and proceeding |
| Proc. of the 6th Int. Workshop on SAME 2013 - Workshop Defining the Research Agenda for Inf. Management and Systems Supporting Sustainable Communities with Smart Media and Automated Systems | conference and proceeding | International Journal of Services, Technology and Management | journal |
| Proceedings - CSERC 2014: Computer Science Education Research Conference | conference and proceeding | Proceedings of the 2013 Information Security Curriculum Development Conference, InfoSec CD 2013 | conference and proceeding |
| Proceedings of 2nd Computer Science Education Research Conference, CSERC 2012 | conference and proceeding | Journal of Global Information Management | journal |
| Proceedings of the 2012 IEEE Latin America Conference on Cloud Computing and Communications, LatinCloud 2012 | conference and proceeding | Proceedings of the 2013 IEEE 63rd Annual Conference International Council for Education Media, ICEM 2013 | conference and proceeding |
| WISMM 2014 - Proceedings of the 1st International Workshop on Internet-Scale Multimedia Management, Workshop of MM 2014 | conference and proceeding | IEEE Transactions on Learning Technologies | journal |
| Human IT | journal | Proceedings - IEEE 14th International Conference on Advanced Learning Technologies, ICALT 2014 | conference and proceeding |
| Journal of Information Systems Education | journal | Nordic Journal of Digital Literacy | journal |
| Open Learning | journal | 2013 International Conference on Computer Graphics, Visualization, Computer Vision, and Game Technology, VisioGame 2013 | conference and proceeding |
| 2014 11th International Conference and Expo on Emerging Technologies for a Smarter World, CEWIT 2014 | conference and proceeding | 26th Bled eConference - eInnovations: Challenges and Impacts for Individuals, Organizations and Society, Proceedings | conference and proceeding |
| EUROMEDIA 2012 - 17th Annual Scientific Conference on Web Technology, New Media Communications and Telematics Theory Methods, Tools and Applications | conference and proceeding | Conference Proceedings - 2013 2nd National Conference on Information Assurance, NCIA 2013 | conference and proceeding |
| ITiCSE-WGR 2013 - Proceedings of the ACM | conference and | ELPUB 2012 - Social Shaping of Digital Publishing: | conference and |
| Conference on Innovation and Technology in Computer Science Education | proceeding | Exploring the Interplay Between Culture and Technology, 16th International Conference on Electronic Publishing | proceeding |
| New Review of Academic Librarianship | journal | Proceedings of the IASTED International Conference on Software Engineering, SE 2012 | conference and proceeding |
| Proceedings of 2013 IEEE International Conference on Teaching, Assessment and Learning for Engineering, TALE 2013 | conference and proceeding | Journal of Information, Communication and Ethics in Society | journal |
| 2013 5th International Conference on Games and Virtual Worlds for Serious Applications, VS-GAMES 2013 | conference and proceeding | Proceedings - VRCAI 2012: 11th ACM SIGGRAPH International Conference on Virtual-Reality Continuum and Its Applications in Industry | conference and proceeding |
| International Journal of Technology Enhanced Learning | journal | Journal of E-Learning and Knowledge Society | journal |
| Australian Journal of Adult Learning | journal | RUSC Universities and Knowledge Society Journal | journal |
| Electronic Government | journal | Journal of Digital Information | journal |
| Proceedings of IDEE 2013: 2nd International Workshop on Interaction Design in Educational Environments - In Conjunction with the 15th Int. Conference on Enterprise Information Systems, ICEIS 2013 | conference and proceeding | 7th European Conference on Information Management and Evaluation, ECIME 2013 | conference and proceeding |
| Government Information Quarterly | journal | UBICOMM 2013 - 7th International Conference on Mobile Ubiquitous Computing, Systems, Services and Technologies | conference and proceeding |
| International Review of Research in Open and Distance Learning | journal | | |